\documentclass[12pt]{iopart}
\usepackage{graphicx,verbatim,amsfonts}

\begin{document}

\title[]{Stability of the $k=3$ Read-Rezayi state in chiral two-dimensional systems with tunable interactions}

\author{D. A. Abanin$^1$, Z. Papi\'c$^2$, Y. Barlas$^3$, and R. N. Bhatt$^{2,1}$}
\address{$^1$ Princeton
 Center for Theoretical Science, Princeton University, Princeton, NJ
 08544} 
\address{$^2$ Department of Electrical Engineering, Princeton University, Princeton, NJ 08544}
\address{$^3$ National High Magnetic Field Laboratory and Department of Physics, Florida State University, FL 32306, USA}


\begin{abstract}

The $k=3$ Read-Rezayi (RR) parafermion quantum Hall state hosts non-Abelian excitations which provide a platform for the universal topological quantum computation. Although the RR state may be realized at the filling factor $\nu=12/5$ in GaAs-based two-dimensional electron systems, the corresponding quantum Hall state is weak and at present nearly impossible to study experimentally. Here we argue that the RR state can alternatively be realized in a class of chiral materials with massless and massive Dirac-like band structure. This family of materials encompasses monolayer and bilayer graphene, as well as topological insulators. 
We show that, compared to GaAs,  these systems provide several important advantages in realizing and studying the RR state. Most importantly, the effective interactions can be tuned {\it in situ} by varying the external magnetic field, and by designing the dielectric environment of the sample. This tunability enables the realization of RR state with controllable energy gaps in different Landau levels. It also allows one to probe the quantum phase transitions to other compressible and incompressible phases.
\end{abstract}

\maketitle

\section{Introduction}

Strongly correlated phases of electrons confined to move in the plane, subject to a perpendicular magnetic field,
have attracted significant attention since the discovery of fractionally quantized Hall conductivity~\cite{tsg}. 
The profound role of topology in this extreme quantum limit leads to the presence of quasiparticles that carry
a fraction of electron charge~\cite{laughlin}, and fractional (Abelian, or possibly non-Abelian) statistics~\cite{laughlin, mr}.
The prospect of excitations possessing non-Abelian statistics has motivated different schemes for topological quantum computation~\cite{tqc}
based on these systems.

These remarkable phenomena occur in the fractional quantum Hall (FQH) regime, when the number of electrons, $N_e$, 
is comparable to the number of magnetic flux quanta $N_\Phi$ through the two-dimension electron system (2DES).
Correlated FQH liquid states appear at certain partial filling $\nu = N_{e}/N_{\Phi}$ of the active Landau level (LL).
In traditional semiconductor heterostructures, the physics of a partially-filled $n=1$ LL differs significantly
from that of $n=0$ LL, due to the node in the single-particle wavefunction~\cite{prange}. As a consequence, the
hierarchy/composite fermion states~\cite{Haldane86, jainbook}, ubiquitous in the lowest Landau level (LLL), are significantly weakened in 
$n=1$ LL, and some of the more exotic states, such as the Read-Rezayi (RR) parafermion states~\cite{rr_parafermion}, 
are likely to be favored. A number of studies have focused on the simplest, $k=2$ non-Abelian member of the RR sequence --
the Moore-Read (MR) ``Pfaffian" state~\cite{mr}. The Pfaffian, or perhaps its particle-hole conjugate version -- the anti-Pfaffian~\cite{antipf}, with the same topological properties, are widely believed to describe the FQH plateau at $\nu=5/2$~\cite{eover4}.
Quasiparticles of the MR and higher RR states obey the non-Abelian statistics~\cite{mr} which is of interest for
topological quantum computation~\cite{tqc}. However, for the purpose of \emph{universal} topological quantum 
computation, MR state is not sufficient, and one must go to a higher, $k=3$ member of the RR sequence (see Ref.~\cite{tqc} and for a general case Ref.~\cite{hormozi}). 
In GaAs systems, experiments~\cite{xia,kumar} have detected only a weak plateau at $\nu=12/5$, which theoretical work~\cite{rr_parafermion} has tentatively 
identified with the particle-hole conjugate of the $k=3$ RR state. This state, however, is very fragile and has been seen only
in a small number of samples, triggering alternative theoretical proposals for its origin~\cite{bs}. Numerical
calculations suggest that the appearance of the RR $k=3$ state is linked to the finite-width of the 2DES in GaAs~\cite{z3_torus, papic_zds},
but little is known about the stability of $k=3$ RR state as the interaction is tuned away from the bare Coulomb point.
For example, in case of $k=2$ (Moore-Read) state, it is possible to construct a 2-body interaction~\cite{ppds}, resulting from
particle-hole symmetrization of the hard-core 3-body repulsion, which yields a ground-state with high overlap with the exact ground state of Coulomb interaction. This model interaction consists only of $V_1$ and $V_3$ Haldane pseudopotentials (to be defined below).
In case of $k=3$ RR state, such an approximate 2-body hard-core interaction has not been constructed, but one could expect that it requires higher order Haldane pseudopotentials (and perhaps some 3-body terms as well). 
 
One of the main disdvantanges of the GaAs-based devices is that their 2DES is buried inside a larger, three-dimensional
structure. This unfortunate fact fixes the effective interactions at values that are often not optimal for some of the most
interesting FQH states, including the RR series. For example, the MR state is found to lie very close to the boundary with
a compressible phase~\cite{Morf98, Rezayi00}. Another problem stems from the strong dielectric screening and finite well-width~\cite{zds} in GaAs, 
which weaken the electron-electron interactions and make FQH states fragile. This has been a major obstacle in the studies of the non-Abelian states,
which could only be observed in ultra-high-mobility samples~\cite{eover4}. 
Thus, it is desirable to find an alternative high-mobility 2DES with strong effective Coulomb interactions that are adjustable in a broad range.

Recently, a new class of such high-mobility 2DES which host chiral excitations with non-trivial Berry phases, has been discovered. 
These ``chiral" materials include graphene and bilayer graphene~\cite{CastroNeto09}, and, more recently, topological insulators~\cite{KaneHasan},
as well as certain quantum wells~\cite{HgTe}. The chiral nature of the quasiparticles gives rise to new electronic properties, 
including an unusual LL sequence, anomalous Hall effect, and suppression of weak localization~\cite{CastroNeto09}.
When these chiral materials are subject to a perpendicular magnetic field, the kinetic energy quenches into discrete Landau levels,
similar to the usual semiconductors with non-chiral carriers. However, due to the chiral band structure~\cite{massivedirac} and the fact that the surface
of these materials is exposed~\cite{tunable}, they offer new possibilities to tune the effective interactions and explore strongly correlated phases. 

In this paper we analyze two practical ways of tuning the effective interactions in chiral 2DES, and the effect this has on the non-Abelian 
FQH states. As a case study, we choose to focus on the $k=3$ RR state. In a stark contrast to GaAs, we find that the chiral 2DESs allow 
for a more robust RR state along with the possibility of its realization in \emph{several} LLs. Additional insights can be obtained by driving transitions between the RR state and the
Abelian hierarchy state, or the compressible stripe phase~\cite{stripe_bubble}. Such transitions can be implemented in chiral and \emph{massive} 2DES by varying 
the external field. Overall, the tunability of the chiral 2DES allows one to explore a larger region of the effective interactions
than has been achieved in GaAs.

\section{Model}  

We consider a family of 2D materials that are characterized by non-trivial Berry phases. One such material is monolayer graphene, a high-mobility atomically thick 2DES~\cite{CastroNeto09}, where recently several FQH states of the type $\nu=m/3$ have been discovered~\cite{graphene_fqhe}.  A closely
related material, bilayer graphene~\cite{CastroNeto09} has similarly high mobility, and exhibits interaction-induced quantum Hall states at integer filling factors at low magnetic fields~\cite{feldman-09np889}. Graphene and its bilayer are characterized by Berry phase $\pi$~\cite{berrypi} (graphene-like) and $2\pi$~\cite{berry2pi} (bilayer-graphene-like with an energy gap), respectively. 

Much of previous theoretical work on chiral materials has been restricted to graphene, exploring the consequences of the four-fold LL degeneracy (valley and spin) that leads to new SU(2) and SU(4)-symmetric fractional states~\cite{multi-component}. Instead, we focus on the high-field limit, neglecting the multicomponent degrees of freedom, and examine the effects originating from the interplay of the Coulomb interaction and band structure. We consider a family of band structures introduced in Ref.~\cite{massivedirac}, which describe a number of high-mobility materials, including graphene with \emph{massive} carriers (mass is generated either spontaneously, or as a result of sublattice symmetry breaking~\cite{CastroNeto09}), topological insulators~\cite{KaneHasan}, bilayer-graphene~\cite{CastroNeto09}, trilayer graphene~\cite{trilayer}, and similar materials. Pristine graphene, which hosts massless Dirac-like fermions, is contained in this model as a particular case. 
\begin{center}
\begin{figure}[t]
\centerline{\includegraphics[scale=0.85]{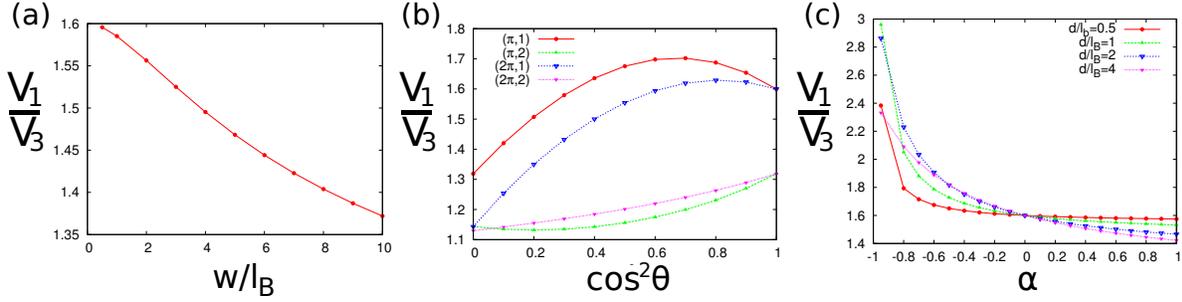}}
\caption[]{(Color online) Ratio $V_1/V_3$ of the Haldane pseudopotentials as a function of system's parameters: the width $w$ in case of the infinite quantum well model (a), the band structure parameter $\theta$ (b), and screening parameter $\alpha$ and distance of the dielectric plate $d$ (c). Chiral 2DES offer the possibility to tune $V_1/V_3$ in a much broader range than it is possible within a quantum well model.}
\label{fig:pp}
\end{figure}
\end{center}
More explicitly, we consider a family of $2\times 2$ Hamiltonians with Berry phase $\pi$ and $2\pi$~\cite{massivedirac}. The case of $\pi$-carriers is realized in graphene, topological insulators, special quantum wells~\cite{HgTe}; the case of $2\pi$-carriers occurs in bilayer graphene, where the energy gap can be controlled in a wide range by a perpendicular electric field~\cite{CastroNeto09}. These $2\times 2$ effective Hamiltonians can be derived pertubatively starting from a tight-binding model for each of the materials with a given chirality $\lambda$, however the effective model will be applicable in a narrower energy range for larger $\lambda$~\cite{yafis}. As it is well known for the case of graphene, the one-body eigenstates of such Hamiltonians possess a spinor structure. Consequently, the harmonic oscillator raising/lowering operators will mix the different components to produce an effective form-factor that has a richer structure than in case of a simple parabolic band such as in GaAs, making it possible to tune the effective interaction.

For $\pi$ carriers, the single-particle wavefunctions are given by~\cite{massivedirac}
$\psi_n=(\cos\theta_n \phi_{|n|-1},\sin\theta_n \phi_{|n|})$, where $\phi_n$ is the wave function of the $n$th non-relativistic LL ($n=\pm 1, \pm 2, \ldots$), and parameter $\theta$ depends on the ratio $\Delta/(\hbar v_0/\ell_B)$, where $\Delta$ is the mass gap, $v_0$ the Fermi velocity, and $\ell_B=\sqrt{\hbar c/eB}$ the magnetic length. We use the notation $(\sigma\pi,n)$ to denote the $n$th LL for $\sigma\pi$ carriers ($\sigma=1,2$).  As a consequence of the spinor wavefunction, the effective form factor~\cite{prange} $F_{n}^{\pi} (q)$ that describes the interaction projected to a $(\pi,n)$ LL is given by
\begin{equation}\label{eq:formfactors}
F_{n}^{\pi} (q)=\cos^2\theta L_{|n|-1} \left(\frac{q^2 \ell_B^2}{2}\right)+\sin^2\theta L_{|n|}\left(\frac{q^2 \ell_B^2}{2}\right),
\end{equation}
where $L_{k}$ is the $k$th Laguerre polynomial, and for simplicity we omitted the index of $\theta$. The form-factor is a mixture of the $(|n|-1)$th and $|n|$th LL form-factors in a non-relativistic 2DES with parabolic dispersion. At  $\theta=\pi/4$, the above equation reduces to the form-factor of graphene~\cite{Nomura06}, however, quite generally by varying $\Delta/(\hbar v_0/\ell_B)$, one can realize any value of $\theta\in (0;\pi/2)$. 

Similarly, for carriers with Berry phase $2\pi$, the single-particle wavefunctions are $\psi_n=(\cos\theta_n \phi_{|n|-1}, \sin\theta_n \phi_{|n|+1} )$, and the form-factor is a mixture of standard $(|n|-1)$th and $(|n|+1)$th form-factors, 
\begin{equation}\label{eq:formfactors2}
F_{n}^{2\pi} (q)=\cos^2\theta L_{|n|-1} \left(\frac{q^2\ell_B^2}{2}\right)+\sin^2\theta L_{|n|+1}\left(\frac{q^2 \ell_B^2}{2}\right). 
\end{equation}

The tunable form of the effective interactions Eqs.(\ref{eq:formfactors},\ref{eq:formfactors2}) provides a way to engineer transitions between strongly correlated phases {\it in situ} by changing the field. As pointed out by Haldane~\cite{prange}, an interacting many-body system of electrons, confined to a partially filled LL, is defined by a finite set of numbers -- the Haldane pseudopotentials. For rotationally invariant systems, these numbers $V_m$ represent the amplitudes of a state of two electrons with the relative angular momentum $m$. The physics of the FQHE is determined by small-$m$ pseudopotentials, i.e. $V_1,V_3,V_5,\ldots$. The role of higher $V_m$'s is to enforce charge neutrality, but they do not affect the incompressibility. Therefore, the ratio of the two strongest pseudopotentials, $V_1/V_3$, can be a rough indicator of the expected many-body phases. In Fig. \ref{fig:pp}(b) we show this ratio for several types of charge carriers and LLs. The variation of $V_1/V_3$ as a function of $\theta$ is to be contrasted with Fig.\ref{fig:pp}(a) where the same quantity is plotted for an infinite GaAs quantum well of width $w$. In GaAs the width of the well can be tuned by electrostatic gates; however, it invariably reduces $V_1/V_3$, whereas the chiral 2DES allow for both increase \emph{and} decrease of $V_1/V_3$, and in a wider range.

Up to this point, we have discussed the change in the effective interaction $V(q)|F(q)|^2$ that resulted from a modification of the single-particle wavefunctions, and therefore $F(q)$. The second method, proposed in Ref.~\cite{tunable}, allows one to directly change $V(q)$. Chiral materials, such as graphene, are often exposed to the environment, which allows for dielectric material to be deposited on top of them. We consider a setup where graphene sample is situated in a dielectric medium with permittivity $\epsilon_1$, and a semi-infinite dielectric plate with permittivity $\epsilon_2\neq \epsilon_1$ is placed at a distance
$d/2$ away from the graphene sheet. The effective interactions between electrons in graphene change due to the surface charges induced at the boundary
between dielectrics~\cite{tunable}:
\begin{equation}\label{eq:interaction_screened}
V(r)=\frac{e^2}{\epsilon_1 r}+\alpha \frac{e^2}{\epsilon_1\sqrt{r^2+d^2}}, \,\, {\rm where} \, \,
\alpha=\frac{\epsilon_1-\epsilon_2}{\epsilon_1+\epsilon_2}.
\end{equation}
The ratio $d/\ell_B$ controls the effective interactions within a partially filled LL. However, the overall energy scale is also modified and this has an impact on the magnitude of the excitation gap. The gap should be multiplied by a factor $\epsilon_{\rm GaAs}/\epsilon_1$ if comparison is to be made with GaAs 2DES. Again, an important advantage of this setup is that the interactions can be tuned {\it in situ} by varying the magnetic field $B$, which modifies the ratio $d/\ell_B$. The consequences for the many-body system are illustrated in Fig.~\ref{fig:pp}(c). In the dielectric case, $V_1/V_3$ is found to have a substantially larger variation as a function of $\alpha$ than the quantum well model or the tuning of the band structure through $\theta$. In particular, for intermediate $d\sim \ell_B$ and negative $\alpha$, $V_1/V_3$ can be readily increased by $50\%$ compared to its bare value ($\alpha=0$).    

Therefore, based on the behavior of $V_1/V_3$, we expect that chiral 2DES offer many possibilities of tuning the interactions. By increasing this ratio, we expect to stabilize the incompressible phases; on the other hand, by reducing it, one can drive a transition to compressible phases, such as stripes and bubbles. In the following, we will explicitly verify these statements in exact diagonalization calculations, and determine the stability of the $k=3$ RR state. In the operational sense, one can define several criteria for the ``stability" of a state; the more of those criteria are met, the more ``stable" the state is. For example, one can compare the overlap (scalar product) between an exact many-body ground state and the RR wavefunction; if the overlap, defined below, is consistenly close to unity for a number of system sizes, we consider it as one of the indicators that the state is in the RR phase. However, there are known examples of different phases of matter which possess ground states of high overlap with each other in finite systems~\cite{gaffnian}; conversely, a relatively poor overlap cannot be taken as a definitive proof that a phase is not realized. 
Additional indicators to identify the phase of matter are, for example, the ground state degeneracy on the torus, and the shift on the sphere (selected to minimize the total ground-state energy). Finally, the robustness of a phase is also determined by the magnitude of the gap for creating neutral and charged excitations. All of these indicators should be assessed together in determining the nature of a phase being studied.  

\section{Method and results}

We consider an interacting, $N$-electron problem with a Landau-level spectrum given above, 
on a compact surface threaded by $N_\Phi$ magnetic flux quanta using exact diagonalization. In the study of FQHE, two kind
of surfaces are available that preserve the translational invariance of an
infinite 2DEG: sphere~\cite{Haldane86} and torus~\cite{pbc}. The two choices of boundary conditions
illustrate the specific features of a FQH state under investigation: on a sphere, the FQH state couples to the curvature of the manifold, which is characterized by the topological number called shift~\cite{shift}. The shift produces a small offset between $N_\Phi$ and the magnetic monopole, whose strength is denoted by $2S$, placed in the center of a sphere. As a consequence, different FQH states ``live" in different Hilbert spaces and in principle can be directly compared only after extrapolation to the thermodynamic limit. On the other hand, the flat surface of a torus leads to a unique definition of $N_\Phi$ for given $N$ and $\nu$, so that different candidate states describing the same filling $\nu$ are all realized in the same Hilbert space. The caveats of this geometry are the additional geometric parameters, the angle and aspect ratio of the torus; the Hamiltonian of a finite-size system depends on these parameters and their specific values favor one FQH phase over the others. Thus, an analysis is slightly more involved but the advantage is that a ground-state degeneracy can be used to identify the phase. For example, in order to detect a stripe phase, we use a rectangular geometry with a specific aspect ratio that best accommodates the stripe in one direction.
In contrast, on the sphere FQH states always appear as non-degenerate, zero angular momentum ground states.

\begin{figure}[t]
\centerline{\includegraphics[scale=0.5]{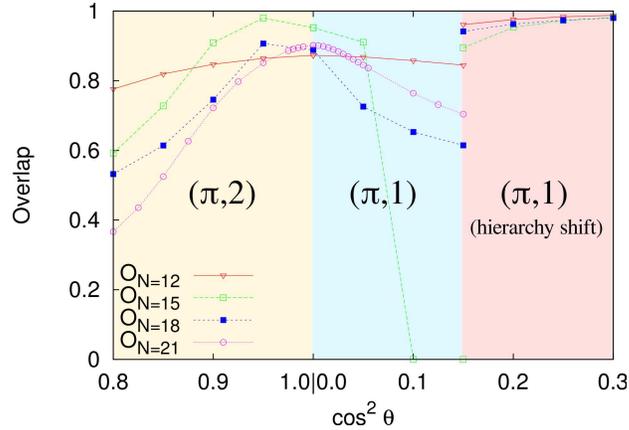}}
\caption[]{(Color online) Overlap between the $k=3$ RR state and the exact ground state in $(\pi,1)$ and $(\pi,2)$ LLs as a function of $\theta$, for systems of different sizes $N$. High overlap is found close to the bare Coulomb interaction in $n=1$ non-relativistic LL. At $\cos^2\theta\approx 0.15$, the ground state undergoes a change in shift, and a hierarchy state sets in (right). For $\cos^2\theta\leq 0.8$ in $(\pi,2)$ LL, the system undergoes an incompressible-compressible transition to a stripe phase.}
\label{fig:z3}
\end{figure}
We will mostly present data for the spherical geometry because this allows for a number of simple diagnostic quantities to be evaluated in a straightforward manner. The spherical representation of broken-symmetry states, however, necessarily contains defects, and periodic boundary conditions have to be used in this case. A number of useful insights can be inferred from the study of the energy spectrum; in addition, we will use overlap calculations to compare an exact, many-body ground state $\Psi_{\rm exact}$, with a numerical representation of a trial wavefunction, $\Psi_{\rm trial}$. The overlap is defined as a scalar product between two normalized vectors, $\mathcal{O}=|\langle \Psi_{\rm trial} | \Psi_{\rm exact} \rangle |$. If $\mathcal{O}$ is consistently close to unity for a number of system sizes considered, we consider the trial wavefunction to be a faithful description of a FQH phase. From the knowledge of a ground-state wavefunction $\Psi_0$, we also evaluate the (projected) static structure factor~\cite{sq}. Sharp peaks in the structure factor indicate the onset of compressible phases~\cite{stripe_bubble_numerics}.

\begin{figure}[t]
\centerline{\includegraphics[scale=0.5]{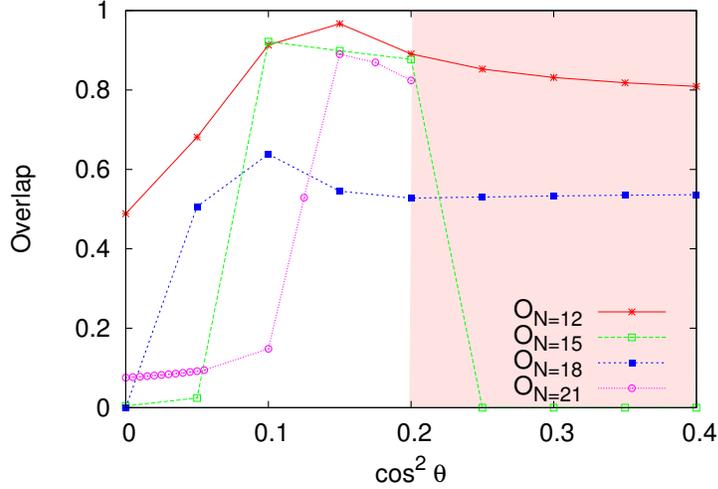}}
\caption[]{(Color online) Overlap between the $k=3$ RR state and the exact ground state in $(2\pi,1)$  LL as a function of $\theta$, for various system sizes $N$. The region of high overlap is found in the window between the stripe phase and the hierarchy phase.}
\label{fig:z3_bilayer}
\end{figure}
In Fig. \ref{fig:z3} we show the overlap between the $k=3$ RR wavefunction and the exact ground state of a system in $(\pi,1)$ and $(\pi,2)$ LLs, as a function of $\theta$. We set the shift is $-3$, corresponding to that of the RR state, and do not consider other candidate states with different values of the shift~\cite{alternative}.
RR model wavefunction can be obtained by diagonalizing a four-body short-range interaction~\cite{rr_parafermion}, which quickly becomes numerically prohibitive as the number of particles grows, or recursively via Jack polynomials~\cite{jack}. The filling factor is fixed at $\nu=3/5$, but in the absence of LL mixing, our results directly apply to $\nu=2/5$ via particle-hole symmetry.
The highest overlap is found for the effective interaction that is in the vicinity of the pure $n=1$ non-relativistic Coulomb potential. Smaller systems (under $N=21$ particles) indicate that RR state might be enhanced for $\cos^2\theta$ slightly less than 1 in $(\pi,2)$ LL, which corresponds to a small relative \emph{decrease} of $V_1$ pseudopotential~\cite{rr_parafermion}. However, the largest system one can obtain by exact diagonalization, $N=21$, suggests that the maximum overlap is practically at the bare Coulomb point. It cannot be determined with certainty whether this is an intrinsic feature of the RR state or a finite size effect that plagues $N=21$ system.  Finite-size effects are generally very strong at this filling factor and further theoretical studies, possibly employing density-matrix renormalization group techniques, are needed. On either side of the bare Coulomb point, the RR state is quickly destroyed. On the left, for $\cos^2\theta\leq 0.8$ in $(\pi,2)$ LL, there is an incompressible-compressible transition to a stripe phase. As mentioned earlier, states with broken translational symmetry are not adequately represented in the spherical geometry; however we can still detect the stripe by the presence of a sharp peak in the structure factor. One arrives at the same conclusion (but more transparently) in the torus geometry, where it is found that the system develops a manifold of nearly degenerate ground states for $\cos^2\theta\leq 0.8$ that are characterized by a linear (uni-directional) array of pseudomomenta that defines the preferred direction of a stripe.  
As we move away from the pure Coulomb point in $(\pi,1)$ LL, the ground state undergoes a change in shift for $\cos^2\theta\geq 0.15$ in order to minimize the energy. The new value of the shift is that of a hierarchy state. 
Therefore, in chiral 2DES, one can in principle study a subtle phase transition between two topological phases with Abelian and non-Abelian excitations.

An interesting feature, shown in Fig. \ref{fig:z3_bilayer}, is the presence of $k=3$ RR correlations in a $(2\pi,1)$ LL. Note that the effective interaction in this case interpolates between pure $n=0$ and $n=2$ (non-relativistic) LL Coulomb repulsion; therefore it is quite different from the $n=1$ LL form-factor (Fig.\ref{fig:z3}). In this case, RR state appears in a narrow window, between the stripe phase (for $\cos^2\theta\leq 0.5$) and the hierarchy state (shaded region in Fig. \ref{fig:z3_bilayer}, for $\cos^2\theta\geq 0.2$). We note that $N=18$ system has a significantly poorer overlap than all the other systems considered, including a larger $N=21$ system, which leads us to conclude that the poor overlap might be a finite-size effect\footnote{Such effects on the sphere often originate from the aliasing problem, but in this case it is not clear what the competing state is. Aliasing problem on the sphere was introduced in N. d'Ambrumenil and R. Morf, Phys. Rev. B {\bf 40}, 6108 (1989).}.
More importantly, the excitation gap in the window where the ground state is well-described by the RR wavefunction is very small (and also smaller than in $(\pi,1)$,$(\pi,2)$ LLs). This, together with the presence of a strong incompressible hierarchy phase, would make it difficult to observe experimentally.

\begin{figure}[t]
\centerline{\includegraphics[scale=0.5]{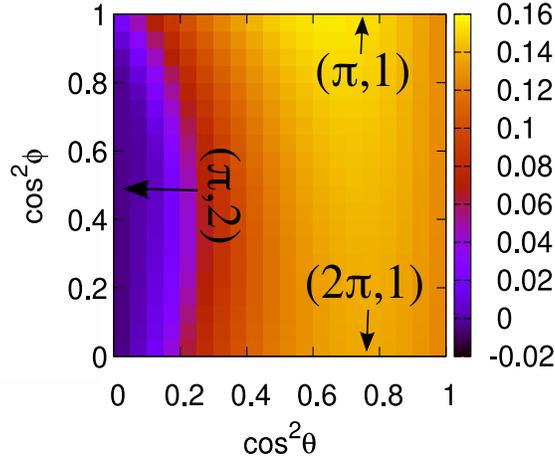}}
\caption[]{(Color online) Charge gap (in units of $e^2/\epsilon \ell_B$) at $\nu=3/5$ for the generalized material described by Eq. \ref{eq:formfactor_general}. The gap smoothly decreases from the hierarchy phase down to zero (stripe phase). RR state appears in the transition region. The gaps are calculated for $N=18$ particles at the shift $-3$.}
\label{fig:chargegap}
\end{figure}
In the remainder of this paper, we investigate the dependence of the excitation gap as we tune the parameters of the system. As mentioned above, the experimental gap at $\nu=12/5$ is tiny. It is therefore essential to find ways to enhance this gap in order to manipulate the state, e.g. via surface probes. To this end, we consider a generalized model for the material that involves a superposition of $n=0$, $n=1$ and $n=2$ LL non-relativistic form factors~\cite{massivedirac}:

\begin{equation}\label{eq:formfactor_general}
F_n(q) = \cos^2\theta L_{|n|} \left(\frac{q^2 \ell_B^2}{2}\right)  +\sin^2\theta \cos^2\phi L_{|n|+1}\left(\frac{q^2 \ell_B^2}{2}\right)  +\sin^2\theta \sin^2\phi L_{|n|+2}\left(\frac{q^2 \ell_B^2}{2}\right).
\end{equation}
 In Fig. \ref{fig:chargegap} we plot the charge gap at $\nu=3/5$ for $N=18$ particles. The ground state of this particular system has a somewhat poorer overlap with the RR state than other system sizes, however the charge gap shows no such difference from other cases.
Along certain lines, indicated by arrows, the phase diagram reduces to one of the cases studied above. Note that proper finite-size scaling needs to be performed in order to get the correct values for the gap; however, it was previously found for filling factors $\nu=1/3$ and $\nu=1/2$~\cite{tunable,massivedirac} that this rigorous analysis produces values that are roughly in agreement with the ones shown in Fig. \ref{fig:chargegap}. At $\nu=3/5$ the number of available system sizes is too small to perform such an extrapolation reliably, and the plot in Fig. \ref{fig:chargegap} should only be viewed qualitatively.  
We find a smooth transition between the hierarchy state and a stripe, with the gaps gradually dropping to zero for small $\cos^2\theta$. 
In the curved, narrow region around the transition, we previously found large overlaps with the RR state~\cite{massivedirac}. The excitation gaps in the transition region are much smaller than in the hierarchy phase, but they are nonetheless higher than at pure $n=1$ non-relativistic LL point ($\cos^2\theta=0$, $\cos^2\phi=1$). 
Therefore, tuning the band structure does allow for a modest enhancement of the gap in the RR phase. Qualitatively similar results are obtained for the neutral, instead of charge, gap.

\begin{figure}[t]
\centerline{\includegraphics[scale=0.45]{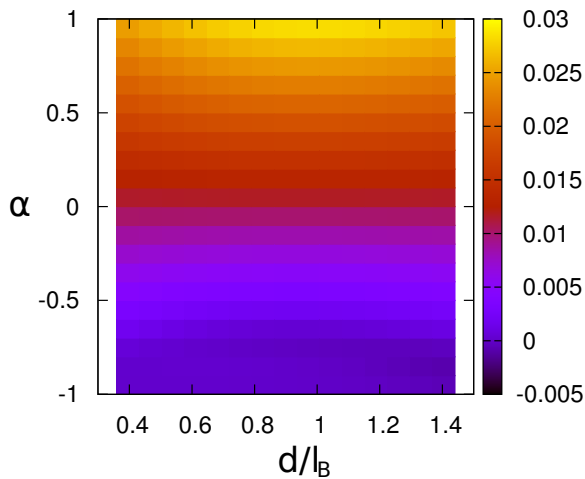}}
\caption[]{(Color online) Charge gap (in units of $e^2/\epsilon_1 \ell_B$) at $\nu=3/5$ as a function of the screening parameter $\alpha$ and the distance of the dielectric, $d/\ell_B$. The gap is largest for $\alpha$ close to unity and intermediate $d\sim \ell_B$. For negative $\alpha$, the gap vanishes and the system is in the stripe phase. Data is shown for $N=18$ particles on the sphere at the shift $-3$.}
\label{fig:dielectric}
\end{figure}

A larger variation of the gap can be achieved in the setup that utilizes the dielectric screening (Eq. \ref{eq:interaction_screened}). In Fig. \ref{fig:dielectric} we show the charge gap as a function of the screening parameter $\alpha$ and the distance of the dielectric plate, $d/\ell_B$. Unless $\alpha$ is close to unity, the gap does not vary significantly with $d/\ell_B$, which hints at the strong finite-size effects present even in this, fairly large, system of $N=18$ particles. However, as $\alpha$ becomes close to unity, the gap displays a clear maximum for $d\sim \ell_B$, which is roughly two times larger than for the bare interaction. The overlap with the RR wavefunction is zero for $\alpha < -0.5$, but fairly large otherwise. Therefore, at least for positive $\alpha$, the gap can be tuned without reducing the good overlap with the model wavefunction.

\section{Conclusion}

In this paper we have explored the stability of $k=3$ RR state, the prime candidate for topological quantum computation, in the phase diagram as we tune the effective interaction via band structure parameters (that depend on the chirality of the material), or via dielectric screening. Numerical calculations suggest that the RR state is most stable in the vicinity of the bare $n=1$ non-relativistic LL Coulomb potential, the same one which describes GaAs materials. Nevertheless, the chiral 2DESs offer several advantages over GaAs. Firstly, we find that RR state can be realized in various (as opposed to a single) LLs, where the interactions have different form from the one in GaAs. More importantly, the excitation gaps can be enhanced and the nature of the RR state can be further probed by driving \emph{in situ} quantum phase transitions, either to compressible states with stripe ordering, or to an incompressible (hierarchy/composite fermion) state. 

\section{Acknowledgements} This work was supported by DOE grant DE-SC$0002140$. Y.B. was supported by the State of Florida. We acknowledge previous collaboration with R. Thomale.


\section*{References}

\begin{thebibliography}{99}

\vspace{-4mm}

\bibitem{tsg}
D. C. Tsui, H. L. Stormer, and A. C. Gossard, Phys. Rev. Lett. {\bf 48}, 1559 (1982).

\bibitem{laughlin}
R. B. Laughlin, Phys. Rev. Lett. {\bf 50}, 1395 (1983).


\bibitem{mr}
G. Moore and N. Read, Nucl. Phys. B {\bf 360}, 362 (1991).



\bibitem{tqc} A. Yu. Kitaev, Ann. of Phys. {\bf 303}, 2 (2003); \emph{ibid} {\bf 321}, 2 (2006);
M. Freedman \emph{et al.}, Comm. Math. Phys. {\bf 227}, 605 (2002); \emph{ibid} {\bf 228}, 177 (2002);
C. Nayak \emph{et al.}, Rev. Mod. Phys. {\bf 80}, 1083 (2008).


\bibitem{prange}
\emph{The Quantum Hall Effect}, 2nd ed., edited by R. E. Prange and S. M. Girvin, Springer-Verlag, New York, 1990.


\bibitem{Haldane86}
F. D. M. Haldane, Phys. Rev. Lett. {\bf 51}, 605 (1983).

\bibitem{jainbook} J. K. Jain, \emph{Composite fermions}, (Cambridge University Press, 2007).



\bibitem{rr_parafermion}
N. Read and E. Rezayi, Phys. Rev. B {\bf 59}, 8084 (1999).

\bibitem{antipf}
M. Levin, B. I. Halperin and B. Rosenow, Phys. Rev. Lett. {\bf 99}, 236806 (2007); S.-S. Lee, S. Ryu,
C. Nayak and M. P. A. Fisher, Phys. Rev. Lett. {\bf 99}, 236807 (2007).

\bibitem{eover4}
R. Willett \emph{et al.}, Phys. Rev. Lett. {\bf 59}, 1776 (1987).

\bibitem{hormozi}
L. Hormozi, N. E. Bonesteel, and S. H. Simon, Phys. Rev. Lett. {\bf 103}, 160501 (2009). 

\bibitem{xia}
J. S. Xia \emph{et al.}, Phys. Rev. Lett. {\bf 93}, 176809 (2004).

\bibitem{kumar}
A. Kumar, G. A. Csathy, M. J. Manfra, L. N. Pfeiffer, and K. W. West, Phys. Rev. Lett. {\bf 105}, 246808 (2010).


\bibitem{bs}
P. Bonderson and J. K. Slingerland, Phys. Rev. B {\bf 78}, 125323 (2008);  Parsa Bonderson, Adrian E. Feiguin, G. M\"{o}oller, J. K. Slingerland, arXiv:0901.4965; A. Wojs, Phys. Rev. B {\bf 80}, 041104(R) (2009).

\bibitem{z3_torus}
E. H. Rezayi and N. Read, Phys. Rev. B {\bf 79}, 075306 (2009).

\bibitem{papic_zds}
Z. Papi\'c, N. Regnault, and S. Das Sarma, Phys. Rev. B {\bf 80}, 201303 (2009). 

\bibitem{ppds}
Michael R. Peterson, Kwon Park, and S. Das Sarma, Phys. Rev. Lett. {\bf 101}, 156803 (2008). 


\bibitem{Morf98}
R. H. Morf, Phys. Rev. Lett. {\bf 80}, 1505 (1998).

\bibitem{Rezayi00}
E. H. Rezayi and F. D. M. Haldane, Phys. Rev. Lett. {\bf 84}, 4685 (2000).

\bibitem{zds} F. C. Zhang and S. Das Sarma, Phys. Rev. B {\bf 33}, 2903
(1986).


\bibitem{CastroNeto09}
A. H. Castro Neto {\it et al.}, Rev. Mod. Phys. {\bf 81}, 109 (2009).

\bibitem{KaneHasan}
M. Z. Hasan and C. L. Kane, Rev. Mod. Phys. {\bf 82}, 3045 (2010). 

\bibitem{HgTe}
B. A. Volkov and O. A. Pankratov, Pisma Zh. Eksp. Teor. Fiz. 42, 145 (1985); JETP Lett. {\bf 42}, 178  (1985); 
B. Buttner {\it et al.}, Nature Phys. {\bf 7}, 418 (2011).


\bibitem{massivedirac}
Z. Papi\'c, D. A. Abanin, Y. Barlas, and R. N. Bhatt, Phys. Rev. B {\bf 84}, 241306 (2011).

\bibitem{tunable}
Z. Papi\'c, R. Thomale, and D. A. Abanin, Phys. Rev. Lett. {\bf 107}, 176602 (2011). 

\bibitem{stripe_bubble}
A. A. Koulakov, M. M. Fogler, and B. I. Shklovskii, Phys. Rev. Lett. {\bf 76}, 499 (1996); 
R. Moessner and J. T. Chalker, Phys. Rev. B {\bf 54}, 5006 (1996).


\bibitem{graphene_fqhe}
X. Du {\it et al.}, Nature {\bf 462}, 192 (2009);
K. Bolotin {\it et al.}, Nature {\bf 462}, 196 (2009);
C. R. Dean {\it et al.}, Nat. Phys. {\bf 7}, 693 (2011). 

\bibitem{berrypi}

K. S. Novoselov \emph{et al.}, Nature {\bf 438}, 197–200 (2005); Y. Zhang \emph{et al.}, Nature {\bf 438}, 201–204 (2005).

\bibitem{berry2pi}
K. Novoselov \emph{et al.}, Nat. Phys. {\bf 2}, 177 - 180 (2006).


\bibitem{multi-component}
V. M. Apalkov and T. Chakraborty, Phys. Rev. Lett. {\bf 97}, 126801 (2006); 
C. T\H{o}ke and J. Jain, Phys. Rev. B {\bf 75}, 245440 (2007); 
Z. Papi\'c, M. O. Goerbig, and N. Regnault, Phys. Rev. Lett. {\bf 105}, 176802 (2010).

\bibitem{trilayer}
W. Bao {\it et al.}, Nat. Phys. {\bf 7}, 948 (2011); T. Taychatanapat {\it et al.},  Nat. Phys. {\bf 7}, 621 (2011); A. Kumar {\it et al.},  Phys. Rev. Lett. {\bf 107}, 126806 (2011); C. H. Lui {\it et al.}, Nat. Phys. {\bf 7}, 944 (2011). 

\bibitem{yafis}
Y. Barlas, K. Yang, and A. H. MacDonald, arXiv:1110.1069.

\bibitem{Nomura06}
K. Nomura and A. H. MacDonald, Phys. Rev. Lett. {\bf 96}, 256602 (2006); M. O. Goerbig, R. Moessner, and B. Dou\c{c}ot,
Phys. Rev. B {\bf 74}, 161407 (2006).

\bibitem{feldman-09np889} B. E. Feldman, J. Martin, A. Yacoby, Nature Phys. {\bf 5}, 889 (2009).


\bibitem{gaffnian}
Steven H. Simon, E. H. Rezayi, N. R. Cooper, and I. Berdnikov, Phys. Rev. B {\bf 75}, 075317 (2007). 

\bibitem{pbc}
D. Yoshioka, B. I. Halperin, and P. A. Lee, Phys. Rev. Lett. {\bf 50}, 1219 (1983); F. D. M. Haldane in Ref.~\cite{prange}.

\bibitem{shift}
X. G. Wen and A. Zee, Phys. Rev. Lett. {\bf 69}, 953 (1992).

\bibitem{alternative}
M. Hermanns, Phys. Rev. Lett. {\bf 104}, 056803 (2010); G. J. Sreejith, C. T\H{o}ke, A. Wojs, and J. K. Jain, Phys. Rev. Lett. {\bf 107}, 086806 (2011).


\bibitem{sq}
S. He, S. H. Simon and B. I. Halperin, Phys. Rev. B {\bf 50}, 1823 (1994).

\bibitem{stripe_bubble_numerics}
E. H. Rezayi, F. D. M. Haldane, and K. Yang, Phys. Rev. Lett. {\bf 83}, 1219 (1999);
F. D. M. Haldane, E. H. Rezayi, and K. Yang, Phys. Rev. Lett. {\bf 85}, 5396 (2000).

\bibitem{jack}
B. Andrei Bernevig and F. D. M. Haldane, Phys. Rev. Lett. {\bf 100}, 246802 (2008).



\end{thebibliography}
\end{document}